\documentclass[aps,prl,twocolumn,showpacs,superscriptaddress,groupedaddress]{revtex4}

\usepackage{graphicx}  
\usepackage{dcolumn}  
\usepackage{bm}        
\usepackage{amsmath}
\makeatletter 
\def\maketag@@@#1{\hbox{\m@th\normalfont\normalsize#1}} 
\makeatother 
\usepackage[toc,page]{appendix}  
\usepackage{graphicx}
\usepackage{amssymb}
\usepackage{amsmath}
\usepackage{algorithm} 
\usepackage[noend]{algpseudocode} 
\usepackage{amsfonts}
\usepackage[mathscr]{eucal}
\usepackage{bbold}
\usepackage{color}
\usepackage{tikz}
\usetikzlibrary{matrix,arrows,decorations.pathmorphing}

\newcommand{\mbf}[1]{\mathbf{#1}}
\newcommand{\ket}[1]{\left| #1 \right\rangle}
\newcommand{\bra}[1]{\left\langle #1 \right|}

\newcommand{\Tr}{\mathrm{Tr}}

\usepackage{soul}

\begin{document}

\title{Optimal Pure-State Qubit Tomography via Sequential Weak Measurements}
\author{Ezad Shojaee}
\affiliation{Center for Quantum Information and Control (CQuIC) \\ 
Department of Physics and Astronomy, University of New Mexico, Albuquerque, New Mexico 87131, USA}
\author{Christopher S. Jackson}
\affiliation{Center for Quantum Information and Control (CQuIC) \\ 
Department of Physics and Astronomy, University of New Mexico, Albuquerque, New Mexico 87131, USA}\author{Carlos A. Riofr{\'i}o}
\affiliation{Dahlem Center for Complex Quantum Systems, Freie Universit{\"a}t Berlin, Berlin, Germany}
\author{Amir Kalev}
\affiliation{Joint Center for Quantum Information and Computer Science (QuICS), University of Maryland, College Park, Maryland 20742, USA} 
\author{Ivan H. Deutsch}
\affiliation{Center for Quantum Information and Control (CQuIC) \\ 
Department of Physics and Astronomy, University of New Mexico, Albuquerque, New Mexico 87131, USA}
                             
\date{\today}

\begin{abstract}
The spin-coherent-state positive-operator-valued-measure (POVM) is a fundamental measurement in quantum science, with applications including tomography, metrology, teleportation, benchmarking, and measurement of Husimi phase space probabilities. We prove that this POVM is achieved by collectively measuring  the spin projection of an ensemble of qubits weakly and isotropically.
We apply this in the context of optimal tomography of pure qubits. We show numerically that through a sequence of weak measurements of random directions of the collective spin component, sampled discretely or in a continuous measurement with random controls, one can approach the optimal bound.
\end{abstract}

\maketitle

In the standard paradigm of quantum tomography, one is given $N$ copies of a quantum state that one seeks to estimate.  When limited only by these finite quantum statistics and no other systematic experimental errors, what is the measurement that achieves the optimal average estimation fidelity?  For the case of qubits, given \textit{a priori} knowledge that the state is pure, this problem was solved long ago in a seminal paper by Massar and Popescu (MP)~\cite{Massar1995}. The optimal average fidelity is $\bar{\mathcal{F}}_{\mathrm{opt}}= (N+1)/(N+2)$, and one can only reach this bound with a measurement that acts {\em collectively} on all $N$ copies. ``Local" measurements acting nonadaptively on one copy at a time can only achieve at best a scaling of $1-\bar{\mathcal{F}} \sim 1/\sqrt{N}$~\cite{Jones1994,Bagan2002, Mahler2013}. 

The MP bound is achieved by a measurement whose positive-operator-valued-measure (POVM) is an overcomplete basis whose elements are proportional to projectors onto spin-coherent states (SCS) of the collective spin $J$ in the symmetric subspace of $N=2J$ qubits. The SCS-POVM is a fundamental measurement in quantum information science, with applications including metrology~\cite{Holevo1982, Appleby2000}, teleportation~\cite{Braunstein2000}, benchmarking~\cite{Yang2014}, and measurement of Husimi phase space probabilities~\cite{Kofler2008}.  While the Glauber-coherent-state-POVM in infinite dimensions has a well-known implementation via heterodyne measurement~\cite{D’Ariano1997}, despite various attempts~\cite{Peres2006quantum, D’Ariano2001, D’Ariano2002}, there is no known implementation of POVMs over generalized-coherent-states for other Lie groups~\cite{Gilmore1972,Perelomov1972}, such as the SU(2)-coherent-states considered here (except for one qubit, $N=1$, $J=1/2$)~\cite{D’Ariano2002}.

The SCS-POVM has been considered physically unattainable and previous works have constructed alternative POVMs that also attain the optimal bound for tomography of qubits and qudits~\cite{Derka1998, Latorre1998, Brub1999, Acin2000, Hayashi2005}.  While in principle one can use the Neumark extension to realize these POVMs consisting of a finite number of measurement outcomes, such constructions have limited applicability, particularly as $N$ grows beyond a few qubits.

In this Letter we show that the SCS-POVM is in fact physically realizable in a direct manner for the application of optimal tomography and other quantum information protocols. In particular, we show that we can realize the SCS-POVM by measuring the collective spin, $\mbf{J} =\sum_{i=1}^N \vec{\sigma}^{(i)}/2$, weakly and isotropically over a sufficiently long time.  This sequence of weak measurements is in a similar spirit to continuous collective measurement tomography~\cite{Silberfarb2005, Riofrio2011, Cook2014}, which has been used for reconstructing states in a fast and robust manner~\cite{Smith2006, Smith2013}  as well as in the ``retrodiction" of initial quantum states~\cite{Gammelmark2013a,Tan2015,Dressel2017,Hacohen-Gourgy2016,Guevara2015}. Here we show that the sequential isotropic protocol asymptotically saturates the MP bound in the appropriate limit.

To establish the foundation and notation, we briefly review the MP bound~\cite{Massar1995}.  We consider $N$ pure qubits all prepared with the same unknown Bloch vector, $\mbf{n}_0$. The $N$-qubit state is $\ket{\Psi_0}\equiv\ket{\uparrow_{\mbf{n}_0}}^{\otimes N} = \ket{J,J}_{\mbf{n}_0}$, a SCS in the $(2J+1)$-dimensional symmetric subspace,  where $\ket{J,M}_{\mbf{n}_0}$ is the Dicke state along $\mbf{n}_0$, $ (\mbf{n}_0\cdot \mbf{J})\ket{J,M}_{\mbf{n}_0} = M\ket{J,M}_{\mbf{n}_0}$.
The SCS form a POVM according to Ref. ~\cite{Peres2006quantum} 
\begin{equation}
\int d\mbf{n}\; E_\mbf{n} =\int d\mbf{n}\; \frac{2J+1}{4\pi} \ket{J,J}_\mbf{n}\bra{J,J}_\mbf{n} = \mathbb{1},
\end{equation}
where $E_\mbf{n}$ denote the POVM elements, proportional to SCS projectors along unit directions $\mbf{n}$, $\int\!d\mbf{n}$ denotes integration over the $4\pi$ steradians of the sphere, and $\mathbb{1}$ is the identity on the symmetric subspace.

If one considers a more general collective POVM, $\{E_r\}$ with outcomes $r$, Banaszek and Devetek have shown~\cite{Banaszek2001} that the state assignment which maximizes the average fidelity 
is $\ket{\uparrow_{\hat{\mbf{n}}_r}}$, where 
\begin{equation}\label{optimal}
\hat{\mbf{n}}_r \equiv \frac{\Tr(E_r \mbf{J})}{\vert \Tr(E_r \mbf{J})\vert}. 
\end{equation}
If $E_r$ is proportional to a SCS along $\hat{\mbf{n}}_r$, this result is consistent with the MP protocol, since $\Tr(E_r \mbf{J}) \propto \bra{J,J}_{\hat{\mbf{n}}_r} \mbf{J}\ket{J,J}_{\hat{\mbf{n}}_r}=\hat{\mbf{n}}_r J$.\\

We show that one can approximate the SCS-POVM to arbitrary precision through a sequence of weak collective measurements.  
The weak measurement of a collective spin component $\mbf{u} \cdot \mbf{J} \equiv J_\mbf{u}$ in the direction $\mbf{u}$ is described by a Kraus operator~\cite{Caves1987},
\begin{equation} \label{diff_Kraus}
\delta K_{\mbf{u}}(m) = \left( \frac{\kappa \delta t}{2 \pi} \right)^{1/4} e^{-\frac{\kappa \delta t}{4} (J_\mbf{u} - m)^{2} },
\end{equation}
where $m$ is a continuous variable outcome, $\kappa$ is the measurement rate, and $\delta t$ is the measurement duration.
Given a state $\ket{\Psi(t)}$, the probability density for outcome $m$ is determined by the Born rule, $P_m(t)=\vert \vert \delta K_m\ket{\Psi(t)} \vert \vert^2$, and $\ket{\Psi(t+\delta t)}=\delta K_m \ket{\Psi(t)}/\sqrt{P_m}$ is the postmeasurement state.
As a result, the weak measurement backaction generally squeezes the uncertainty along the measured direction and gives the mean spin a random kick.

If the direction $\mbf{u}$ is fixed, then the measurement will continually squeeze the uncertainty, ultimately leading to a projective measurement onto an eigenstate of $J_\mbf{u}$. On the other hand, if we consider a collection of the directions $\{\mbf{u}_i\}$ that are chosen isotropically, and each measurement is sufficiently weak such that $\kappa \delta t \ll \Delta J^2_{\mbf{u}_i}$, then we expect the effect of squeezing to ``average out" and the state to remain close to a SCS~\cite{Cook2014}. Thus, the net effect of the measurement backaction will be a random walk of the mean spin on the sphere. After some time the postmeasurement state will have diffused sufficiently far from the initial state, a distance of order $\sqrt{N}$, and no further information about the initial state will remain. The maximum fidelity is limited, thus, by the total number of copies due to the measurement backaction.

With this physical intuition, we specify our protocol for approaching the MP bound with a physically implementable unraveling of the SCS-POVM. Consider a sequence of weak measurements along the $L$ directions, $\{\mbf{u}_1, \mbf{u}_2,\dots, \mbf{u}_L\}$. A measurement record $r\equiv\{m_1,m_2,\dots,m_L\}$ defines a total effect specified by the POVM element $E_r=K_r^\dag K_r$, where the total Kraus operator is $K_r=\prod_{i=1}^L \delta K_i$, with $\delta K_i = \delta K_{m_i}(\mbf{u}_i)$ given in Eq. (\ref{diff_Kraus}). Operators in an indexed product are understood here as ordered from right to left.

In order to achieve a SCS-POVM, one must be able to remove the effects of squeezing due to the quadratic operators $J_{\mbf{u}_i}^2$ therein. 
This can be done by grouping together $l$ weak measurements into time intervals $\Delta t = l \delta t$. For the $I$th interval, $l_I \equiv  \{(I-1)l+1, \ldots, Il\}$, the resulting Kraus operator is
\begin{eqnarray}\label{renorm}
\Delta K_I &\equiv& \prod_{i\in l_I}  \delta K_i \propto \prod_{i\in l_I} e^{-\frac{\kappa \delta t}{4}  J^2_{\mbf{u}_i}} e^{\frac{\kappa \delta t}{2} m_i J_{\mbf{u}_i}}\\
&=& \exp{\left\{ -\frac{\kappa\delta t}{4}\sum_{i\in l_I} J^2_{\mbf{u}_i}+ \frac{\kappa \delta t}{2}  \sum_{i\in l_I}  m_i J_{\mbf{u}_i} + \ldots \right\} }, \nonumber
\end{eqnarray}
as follows from the Baker-Campbell-Hausdorff expansion. If the $l$ measurements are isotropic, then
\begin{equation}\label{iso}
\frac{1}{l}\sum_{i\in l_I}  J^2_{\mbf{u}_i} = \mbf{J}\cdot \left(\frac{1}{l}\sum_{i\in l_I}  \mbf{u}_i \mbf{u}_i \right)\cdot \mbf{J} = \frac{1}{3} \mbf{J}^2.
\end{equation}
Thus, for sufficiently weak measurements such that $\kappa\Delta t \ll 1$,
the quadratic squeezing terms average out because $\mbf{J}^2= J(J+1)\mathbb{1}$ is proportional to the identity.

Let us define the ``operator valued'' part $\tilde{K}(t)$ of the total Kraus operator such that
\begin{equation}
K_r(L \delta t) = \left(
\frac{\kappa \Delta t}{2\pi}\right)^{\!L/4 l}
\!\!\exp\left(-\frac{\kappa\Delta t}{4}\!\sum_{I=1}^{L/l}\!\boldsymbol{\mu}_I^2\right)e^{-\frac{\kappa t}{12}\mbf{J}^2}\tilde{K}_{\boldsymbol{\mu}}(t).
\end{equation}
In the limit $\kappa\Delta t \ll 1$, $\tilde{K}_{\boldsymbol{\mu}}(t)$ is the solution to the differential equation
\begin{equation}\label{geneq}
\frac{d}{dt}\tilde{K}(t) = \frac{\kappa}{2} \boldsymbol{\mu}(t)\!\cdot\!\mbf{J} \tilde{K}(t)
\end{equation}
with initial condition $\tilde{K}(0)=\mathbb{1}$.
The collection of these operator values enumerated by the coarse-grained measurement records $\boldsymbol{\mu}(t)$ define a completely positive superoperator 
\begin{equation}\label{CPOVPF}
\mathcal{Z}_t(\rho) = \int\!\!\mathcal{D}\boldsymbol{\mu}\,\tilde{K}_{\boldsymbol{\mu}}(t)\rho\tilde{K}_{\boldsymbol{\mu}}(t)^\dag
\end{equation}
where we have defined the Wiener measure
\begin{equation}\label{meas}
\mathcal{D}\boldsymbol{\mu} = \left(\frac{\kappa \Delta t}{2\pi}\right)^{\frac{L}{2 l}}e^{-\frac{\kappa\Delta t}{2}\!\sum_{I=1}^{L/l}\!\boldsymbol{\mu}_I^2}\prod_{I=1}^{L/l}d\boldsymbol{\mu}_I.
\end{equation}
Given this Gaussian form, we see that the operator values $\tilde{K}_{\boldsymbol{\mu}}$ are elements in an ensemble of paths generated by an isotropic Wiener process.
Since the measure is isotropic for each $\boldsymbol{\mu}_{I}$ the resulting POVM will be rotationally invariant, as expected.

Significantly, the commutators of the generators in Eq. (\ref{geneq}) are in the six-dimensional span of $\{-iJ_k,J_k\}$ which is a representation of the Lie algebra $\mathfrak{sl}(2,\mathbb{C})$. Therefore each $\tilde{K}_{\boldsymbol{\mu}}$ at every time step is proportional to the representation of a member of the Lie group $\mathrm{SL}(2,\mathbb{C})$, rather than the entire $\mathrm{SL}(2J+1,\mathbb{C})$, which would be generated if generators such as $J_{\mbf{u}}^2$ were present in the differential equation.
Such operators can be decomposed into  a restricted polar form,
\begin{equation}\label{polar}
	\tilde{K}(t) = U(t) e^{\boldsymbol{\alpha}(t)\cdot\mbf{J}},
\end{equation}
where $U(t)$ is a representation of an element of $\mathrm{SU}(2)$, $\boldsymbol{\alpha}(t) = \alpha(t)\mbf{n}(t)$ is real, and $\mbf{n}(t)$ is a unit vector. It follows that the operator-valued part of the POVM element corresponding to the sequence of weak measurements, sampled isotropically over the sphere after a time $T$ is
\begin{eqnarray}\label{E(T)}
\tilde{E}(T) &=& \tilde{K}^\dag(T)  \tilde{K}(T) = e^{2\boldsymbol{\alpha}(T)\cdot \mbf{J}} \\
&= & \sum_{M=-J}^{J}  e^{2\alpha(T) M} \ket{J,M}_{\mbf{n}(T)}\bra{J,M}_{\mbf{n}(T)}. \nonumber
\end{eqnarray}
We will show that $\alpha(T)$ has a variance which increases diffusively with time, $\overline{\alpha(T)^2} \propto \kappa T$.
This implies that for $\kappa T \gg 1$, the probability that $|\alpha(T)|<A$  decreases asymptotically in time as $A/\sqrt{\kappa T}$, and thus in the long time limit, only projectors of highest $M=\pm J$ are  statistically significant instances of the superoperator of Eq. (\ref{CPOVPF}). Thus, each POVM element converges to $\tilde{E}(T) = e^{2\vert \alpha(T) \vert J} \ket{J,\pm J}_{\pm \hat{\mbf{ n}}_{r}}\bra{J,\pm J}_{\pm \hat{\mbf{n}}_{r}}$, proportional to a SCS projector along an asymptotically constant direction $\hat{\mbf{n}}_{r} =\pm \lim_{T\rightarrow \infty} \mbf{n}(T)$.  Together with the rotation-invariant property, one can thus conclude that the sequential weak isotropic measurement protocol realizes the SCS-POVM.

To prove this, write the polar decomposition as
\begin{equation}\label{svd}
\tilde{K}(t) = U(t) V(t) e^{\alpha(t) J_z} V^\dag(t),
\end{equation}
where $J_{\mbf{n}(t)} =V(t) J_z V^\dag(t)$. We define the generator of this unitary map as $\frac{d}{d t}V = -i(\mbf{A}(t)\cdot \mbf{J} ) V(t)$, where $\mbf{A}(t)$ is a real vector that we choose to satisfy $\mbf{n}(t)\cdot \mbf{A}(t) =0$ for convenience. It then follows that
\begin{eqnarray}\label{dsvd}
\frac{d \tilde{K}}{d t} &= &\left[ \frac{d U}{dt} U^\dag + \frac{d\alpha}{dt}\; U \; \mbf{n}\cdot \mbf{J}\; U^\dag \right. \\
&+& \left. i U \left( e^{\boldsymbol{\alpha} \cdot \mbf{J} }\; \mbf{A}\cdot \mbf{J} \; e^{-\boldsymbol{\alpha}\cdot \mbf{J}}  - \mbf{A}\cdot \mbf{J} \right) U^\dag \right] \tilde{K}. \nonumber
\end{eqnarray}
For a rotation by an imaginary angle,
\begin{equation}
 e^{\boldsymbol{\alpha} \cdot \mbf{J}}\; \mbf{A}\cdot \mbf{J}\; e^{-\boldsymbol{\alpha}\cdot \mbf{J}}
 = \cosh \alpha \;\mbf{A}\cdot \mbf{J} + i \sinh \alpha \; (\mbf{n}\times \mbf{A}) \cdot \mbf{J}.
\end{equation}
Comparing Eq. (\ref{geneq}) to Eq. (\ref{dsvd}) and taking the Hermitian part,
\begin{equation}
\frac{d\alpha}{dt}\; \mbf{n}\cdot \mbf{J} - \sinh \alpha \; (\mbf{n}\times \mbf{A}) \cdot \mbf{J} = \frac{\kappa}{2} \left({R} \boldsymbol{\mu}\right) \cdot \mbf{J} ,
\end{equation}
where we define $U^{-1} ( \boldsymbol{\mu}\cdot \mbf{J}) U \equiv ({R} \boldsymbol{\mu})\cdot \mbf{J} $.
Equating the components orthogonal and parallel to $\mbf{n}$,
\begin{eqnarray}\label{svmag}
\frac{d \alpha}{dt} &=& \frac{\kappa}{2} \; \mbf{n}\cdot(R\boldsymbol{\mu}), \\
\label{svdir}\mbf{A} &=& \frac{\kappa}{2\sinh\alpha} \mbf{n} \times ( {R} \boldsymbol{\mu}).
\end{eqnarray}
Integrating Eq. (\ref{svmag})
\begin{equation}
\alpha(T) = \frac{\kappa}{2} \int_0^T  dt \; \mbf{n}(t)\cdot \big({R}(t)\boldsymbol{\mu}(t)\big).
\end{equation}
By Eq. (\ref{meas}), the $\boldsymbol{\mu}(t)$ are isotropically Gaussian distributed, and thus the variables in the integrand $\mbf{n}\cdot \left(R\boldsymbol{\mu}\right)$ are Gaussian distributed with the same (time-independent) variance.
It follows that $\overline{\alpha(T)^2}=\frac{1}{12}{\kappa T}$ increases diffusively with the number of isotropic weak measurements, where $\overline{f[\boldsymbol{\mu}]}=\int\!\mathcal{D}\boldsymbol{\mu}f[\boldsymbol{\mu}]$.

This growth of $\alpha(T)$ implies that every statistically significant element of the Kraus ensemble is proportional to an operator of the form $U(T)\ket{J,J}_{\hat{\mbf{n}}_{r}}\bra{J,J}_{\hat{\mbf{n}}_{r}}$.
Specifically, according to Eq. (\ref{svdir}), as $\alpha(T) \rightarrow \infty$, so must $\mbf{A}(T) \rightarrow 0$ and thus $dV/dt \rightarrow 0$. 
This means that $V$ becomes asymptotically constant and thus $\pm$$\mbf{n}(T) \rightarrow \hat{\mbf{n}}_r$.
Therefore, the direction of the SCS POVM element converges to an estimate of the initial qubit direction. 

Let us further define $\frac{d}{dt}U = -i(\mbf{B}\cdot\mbf{J})U$. Comparing the anti-Hermitian parts of Eq. (\ref{geneq}) and Eq. (\ref{dsvd}), and substituting Eq. (\ref{svdir}) into the result one finds,
\begin{equation}\label{svpost}
	\mbf{B} = \frac{\kappa(\cosh\alpha-1)}{2\sinh\alpha} (R^{-1}\mbf{n}) \times \boldsymbol{\mu}.
\end{equation}
As $\alpha(T) \rightarrow \infty$, $\mbf{B}(T)$ becomes constant in magnitude and thus $U(T)$ wanders perpetually.  This implies that in any realization of a sequence of weak measurements, the postmeasurement state continues to diffuse over the sphere for all times, as expected. 

Any physical realization of this measurement protocol will differ from the idealized model in a number of fundamental respects.  First, each measurement will have a finite duration $\delta t$.   Second, if we choose the $l$ directions as a random sampling of measurements over the sphere, it will be only approximately isotropic. Finally the idealized measurement will be corrupted by decoherence at a rate $\gamma$. Throughout we assume $\kappa \gg \gamma$ and ignore decoherence in the simulations below.  

As an example of a physical realization, consider tomography on atomic spins via continuous measurement as studied in Refs.~\cite{Silberfarb2005, Smith2006, Riofrio2011,Cook2014,Smith2013}.  Using the Faraday interaction and polarization spectroscopy, one can perform a collective $J_z$ measurement of the spins when the laser probe couples uniformly to the atomic ensemble (here $z$ is the propagation direction of the probe)~\cite{Smith2004}.  The measurement rate is $\kappa = C \gamma_s$, where $\gamma_s$ is the photon scattering rate and $C$ is the cooperativity per atom.  The measurement will be weak when the duration of the probe pulse $\delta t \ll 1/\kappa$; decoherence is negligible if $C \gg 1/(\gamma_s T)$. For example, the requisite strong atom-light interface has been demonstrated for $> 40$ atoms in an optical fiber cavity, with observed $C \sim 100$~\cite{Haas2014}.  In such a geometry, one could perform a QND measurement sequence that is decoherence-free to good approximation in a time $T \sim 1/\kappa$.   Finally, to measure an arbitrary spin projection $J_{\mbf{u}_i}$ one can precede the $J_z$ measurement with a physical rotation of the atomic spin direction $\mbf{u}_i \rightarrow \mbf{z}$. 

To demonstrate how one attains the optimal measurement we have performed two types of numerical simulations: (i) sequential random weak measurements; (ii) continuous weak measurements concurrent with time-dependent Hamiltonian control. In type (i), we consider a set of measurement directions $\{\mbf{u}_1,\mbf{u}_2,\dots,\mbf{u}_L\}$ randomly sampled on the sphere by the Haar measure. We simulate random measurement outcomes $m_i$ sampled from the probability distribution $P_i(m_{i}) = \bra{\Psi_{i-1}} \delta K^{\dagger}_i \delta K_i \ket{\Psi_{i-1}}$ using Monte Carlo simulations. The postmeasurement state is determined by $\ket{\Psi_{i}} = \frac{\delta K_{i} \ket{\Psi_{i-1}}}{\vert\vert \delta K_{i} \ket{\Psi_{i-1}}\vert \vert}$, which forms the input that determines the probability distribution for the next measurement outcome, $m_{i+1}$, and the procedure is iterated for $L$ outcomes.  In our simulations we choose $\kappa \delta t = 10^{-4}$.

\begin{figure}
\hspace*{-0.5cm} 
\includegraphics[scale=0.28]{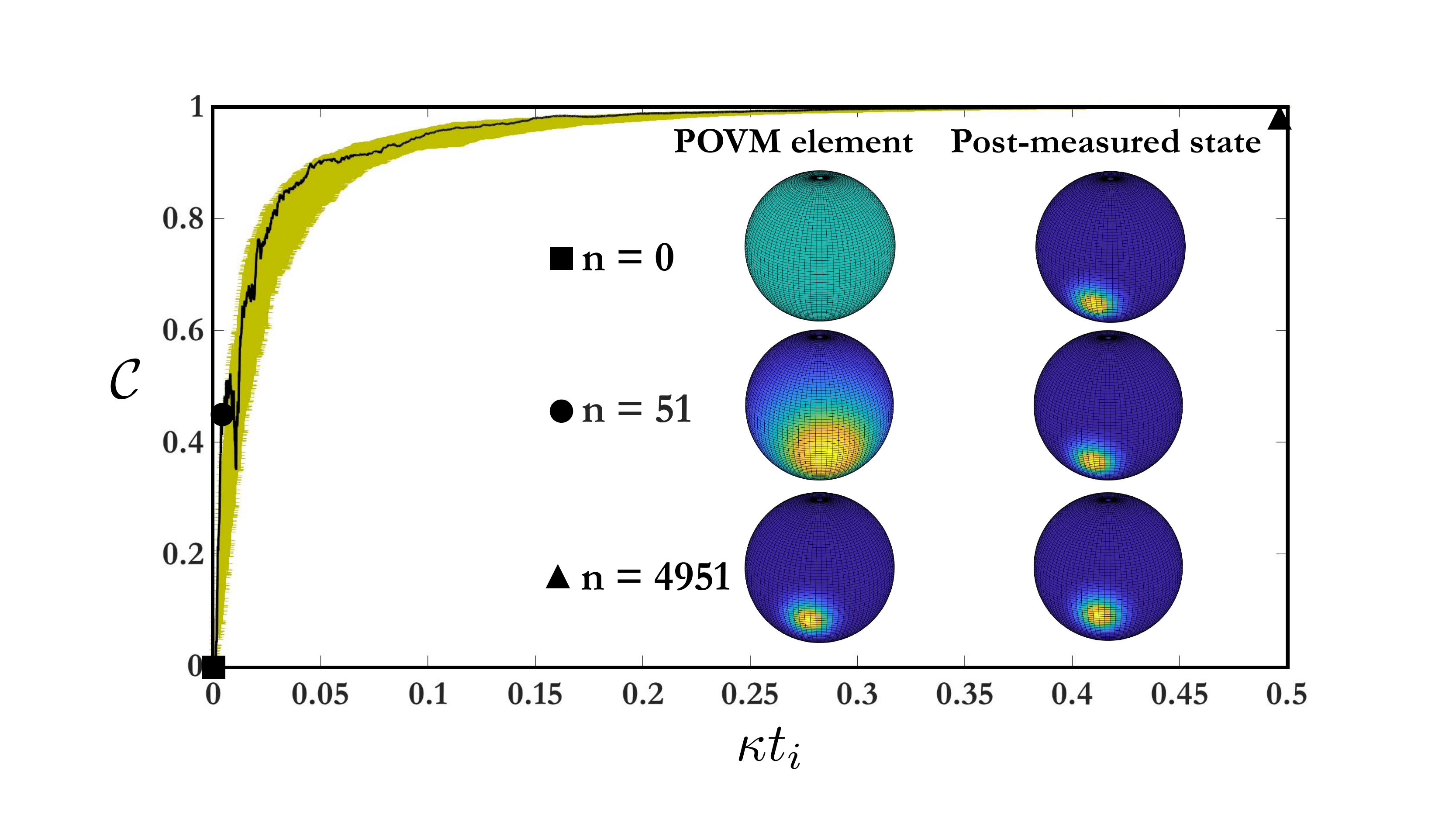}
\caption{\label{fig:Coherency}Simulation of quantum state estimation of a random state of 50 qubits. The coherency parameter of the POVM element, Eq. (\ref{coherency}), is shown as a function of time $t_{i}$ after the $i$th measurement. The parameter converges to $1$, corresponding to a rank-1 operator proportional to a SCS. The green region consists of 50 different realizations; the solid black line is a particular instance. The inset demonstrates the evolution of the Husimi distribution of the POVM element and the postmeasurement state at distinct time steps for this realization.}
\label{Coherency}
\end{figure}

For a given simulated measurement record, $r=\{m_1,\dots,m_L\}$, the POVM element is $E_r = K^\dag_r K_r$ where $K_r = \prod_i^L \delta K_i$.  We can test to see how this converges through the ``coherency parameter'' which satisfies the inequality
\begin{equation}\label{coherency}
\mathcal{C} \equiv \frac{\vert \text{Tr} (\mbf{J} E_{r})\vert^{2}}{J^{2} (\text{Tr}E_{r})^{2}} \leq 1
\end{equation}
for any positive operator $E_r$. The upper bound is achieved {\em iff} $E_{r}$ is a rank-1 operator, proportional to a SCS projector.  Figure \ref{Coherency} shows $\mathcal{C}(t)$, for $0\le t \le 0.5/\kappa$, i.e., 5000 random directions, for $N=50$ copies of the qubit, and 50 different simulated measurement records of a given initial SCS.  We see that $\mathcal{C}$ quickly converges to one for all realizations.   The simulation also shows the expected diffusion of the postmeasurement state over longer times, once the POVM element converges.  

The time constant for the POVM to converge will depend on the number of copies qubit $N$. As new information is gained, we gain finer resolution of the spin direction.  Eventually, the resolution will be better than the spin projection uncertainty $~\sim \sqrt{N}$ and measurement backaction will erase the initial condition.  If the measurement direction is fixed, the resolution $\sim 1/\kappa T$, and we expect the time at which backaction becomes nonnegligible to scale as $\kappa T =\mathcal{O}(1/N)$.  Here, for an isotropic measurement, we can use the coherency parameter to set a timescale for measurement backaction and convergence of the POVM.  We expect from Eq. (\ref{E(T)}) that $E_{r} \propto e^{2 \alpha(T) J_{\mbf{n}(T)}}$, and thus
\begin{equation}\label{dist}
\begin{aligned} 
\mathcal{C}(T) &= \left( 1 - \tfrac{1}{J} \sum\limits_{m=1}^{N} e^{-m 2\alpha(T)} \frac{1-e^{-(N+1-m) 2\alpha(T)}}{1-e^{-(N+1) 2\alpha(T)}} \right)^{2} \\
&\xrightarrow{\alpha(T)\gg 1} 1 -2[ e^{ 2\alpha(T)}(N+1)]^{-1}.
\end{aligned}
\end{equation}
In this case we see that the POVM converges when $e^{2\alpha(T)} \gg 1/(N+1)$, which depends on the diffusive growth of $\alpha(T)$, or $\kappa T =\mathcal{O}(\mathtt{polylog}(1/N))$.

\begin{figure}
\hspace*{-0.5cm} 
\includegraphics[width=1.1 \linewidth, height=5cm] {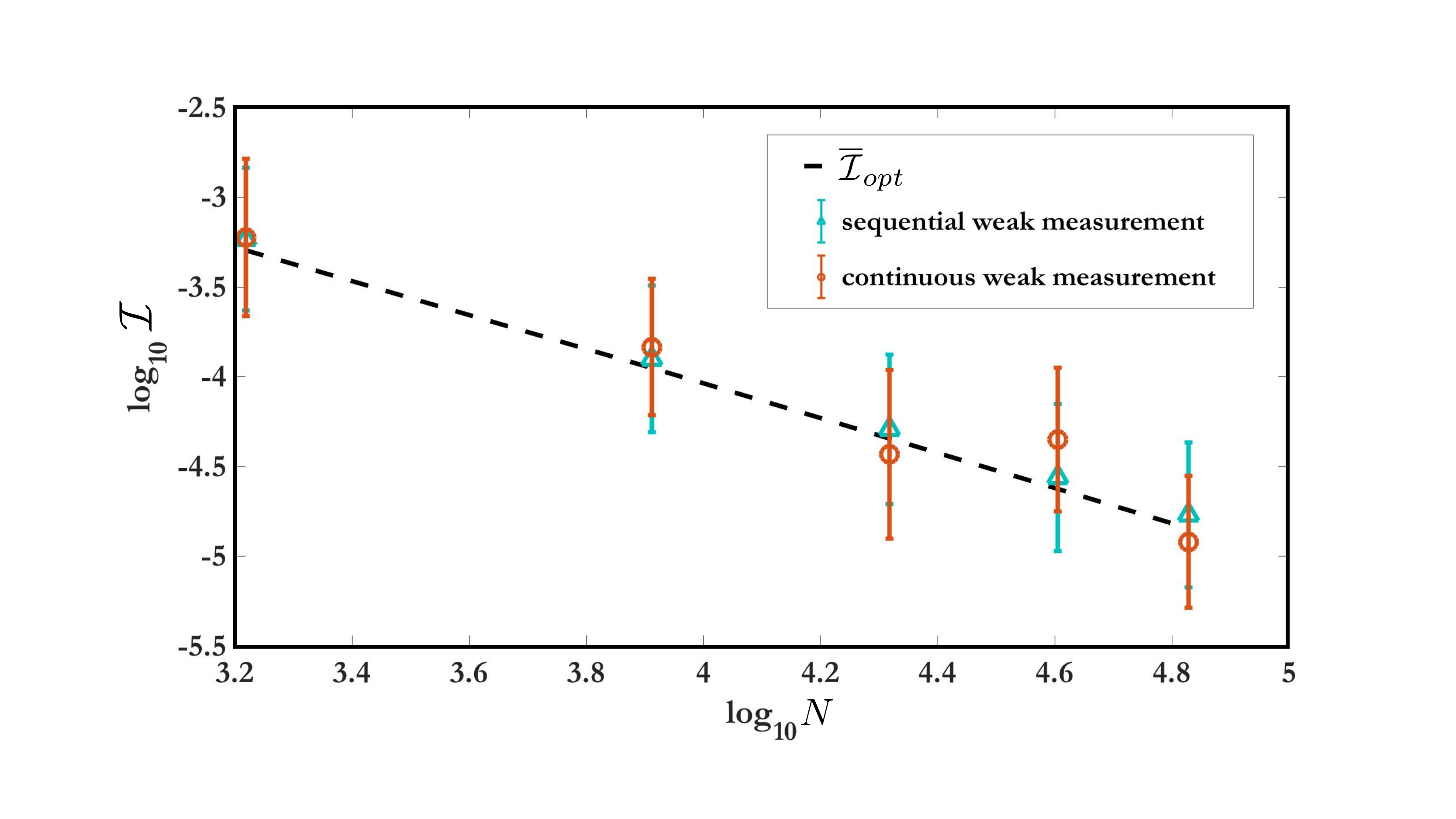}
\caption{\label{fig:fidelity} Average infidelity of reconstruction as a function of the number of qubits for a sequence of weak measurements along random directions (cyan, triangle) and weak measurement along a direction continuously changing in time (orange, circle); see text for parameters. The average and error bars are shown for 400 random directions of the initial SCS.}
\end{figure}

We also test how well this measurement protocol achieves the MP bound by using the simulated record to estimate the initial state according to Eq. (\ref{optimal}). Figure \ref{fig:fidelity} shows the simulated infidelity  $\mathcal{I}=1-\mathcal{F}$ averaged over 400 Haar random initial SCS as a function of $N$. The total measurement time is taken to be $T=1/\kappa$ in all cases.  The MP bound $\bar{\mathcal{I}}_{opt}=(N+2)^{-1}$ is shown for comparison. The simulation is consistent with near optimal tomography.

In type (ii) we simulate continuous weak measurement while simultaneously subjecting the system to a time-dependent external control~\cite{Silberfarb2005, Smith2006, Riofrio2011, Smith2013, Cook2014}.  In this case the measurements occur in infinitesimal time intervals, and random controls can be used to  sample random directions on the sphere, but there are correlations between measurement directions for short times, contrary to the idealizations of our proof. The state evolves according to the stochastic Schr{\"o}dinger equation $\ket{\Psi(t+dt)}=dK(t+dt,t) \ket{\Psi (t)}$, where the differential Kraus operator is 
\begin{equation}
dK(t,t+dt) = \mathbb{1} -iH(t)dt - \tfrac{1}{8}\kappa J_{z}^{2} dt + \tfrac{\sqrt{\kappa}}{2} J_{z}dy(t)
\end{equation}
and
\begin{equation}
dy(t) = \sqrt{\kappa} \bra{\Psi(t)} J_{z}\ket{\Psi(t)} dt + dW(t)
\end{equation}
is the differential measurement record [$dW(t)$ is the Wiener increment]~\cite{Jacobs2006, Gammelmark2013}. 

We simulate the evolution by updating the state with this differential Kraus operator for time increments such that $\kappa dt=10^{-3}/(8J)$. The control Hamiltonian is taken to be $H(t) = \Omega [ \cos\phi(t) J_x + \sin\phi(t) J_y ]$ with $\Omega / 2\pi = 10 \kappa$; $\phi(t)$ is the angle of a time-dependent magnetic field in the $x$-$y$ plane. We choose $\phi(t)$ to be piecewise constant so the spins precess about a magnetic field that has a fixed amplitude but a random direction in the equator that changes every  $\tau = 1/(50 \kappa)$. Such a control policy is sufficient to achieve an informationally complete measurement record \cite{Cook2014}.  

Given a measurement record, we estimate the initial  Bloch vector of a qubit in the atomic ensemble, using Eq. (\ref{optimal}), with $E_r = K^\dag_r K_r$ and $K_r = \prod_{i=1}^{T/dt} dK(t_i+dt,t_i)$. Figure~\ref{fig:fidelity} shows how the continuous measurement performs compared to our random sequential weak protocol and the MP bound. 

In summary, we have shown that one can implement a POVM whose outcomes are specified by the overcomplete set of spin-coherent states via a sequence of weak measurements that are isotropic over the sphere.  The SCS-POVM allows for optimal tomography of pure qubits, metrology, and other applications. The mathematical proof and techniques we have developed are generalizable to qudits, continuous variable systems, and other generalized-coherent-state POVMs of an arbitrary compact semisimple Lie group~\cite{Jackson2018}.  Of particular interest is the possibility of a generalized weak measurement protocol to measure the initial $k$-body correlation functions in a symmetric ensemble.

We thank Alexandre Korotkov and Hendra Nurdin for helpful discussions and insights. This work was supported by the National Science Foundation under Grants No. PHY-1606989 and No. PHY-1630114.  A.K. acknowledges support from the U.S. Department of Defense. C.A.R. was supported by the Freie Universit{\"a}t Berlin within the Excellence Initiative of the German Research Foundation.

\bibliographystyle{apsrev}
\bibliography{references_new.bib} 

\end{document}